\documentclass[aps,prl,twocolumn,10pt,superscriptaddress,longbibliography]{revtex4-2}

\usepackage[utf8]{inputenc}
\usepackage[T1]{fontenc}
\usepackage{amsmath,amsfonts,amssymb}
\usepackage{graphicx}
\usepackage{bm}
\usepackage{hyperref}
\usepackage{microtype}
\usepackage{booktabs}
\usepackage{xcolor}

\newcommand{\lsno}{La$_2$SmNi$_2$O$_7$}
\newcommand{\lno}{La$_3$Ni$_2$O$_7$}

\newcommand{\hz}{h_{z^2}}
\newcommand{\hx}{h_{x^2-y^2}}

\begin{document}

\title{Direct imaging of the Ni $e_g$ hole orbitals in the bilayer nickelate \lsno}

\author{Ethan Saïman}
\affiliation{Université Paris-Saclay, CNRS, Laboratoire de Physique des Solides, 91405 Orsay, France}
\affiliation{Synchrotron SOLEIL, L'Orme des Merisiers, Saint Aubin BP 48, 91192 Gif-sur-Yvette, France}

\author{Feiyu Li}
\affiliation{State Key Laboratory of Crystal Materials, Institute of Crystal Materials, Shandong University, Jinan, Shandong 250100, China}
\affiliation{College of Materials Science and Engineering, Qingdao University of Science \& Technology, Qingdao, Shandong 266042, China}

\author{Junjie Zhang}
\affiliation{State Key Laboratory of Crystal Materials, Institute of Crystal Materials, Shandong University, Jinan, Shandong 250100, China}

\author{Jean-Pascal Rueff}
\affiliation{Synchrotron SOLEIL, L'Orme des Merisiers, Saint Aubin BP 48, 91192 Gif-sur-Yvette, France}
\affiliation{Sorbonne Université, CNRS, Laboratoire de Chimie Physique – Matière et Rayonnement, LCPMR, F-75005 Paris, France}

\author{Victor Balédent}
\email{victor.baledent@universite-paris-saclay.fr}
\affiliation{Université Paris-Saclay, CNRS, Laboratoire de Physique des Solides, 91405 Orsay, France}
\affiliation{Institut universitaire de France (IUF)}

\date{\today}

\begin{abstract}
Models of unconventional superconductivity in cuprates and nickelates depend sensitively on the distribution of valence electrons among the $d$ orbitals. But accurately determining the orbital electron or hole occupations in these systems remains a difficult and unresolved task. Here, using $s$-core-level non-resonant inelastic x-ray scattering at the Ni $M_1$ edge, we determine the angular distribution of the Ni $3d$ holes in a single crystal of \lsno. Our results demonstrate that the hole occupation ratio $\hz/\hx = 0.94\pm0.06$, thus significantly above the expected value of $2/3$ for a filled $d_{z^2}$ orbitals models and close to value expected in a Hund's-coupling picture. Our results support superconductivity scenarios in which both $e_g$ orbitals contribute to the electron-pairing mechanism in \lsno.
\end{abstract}

\maketitle

Superconductivity in transition-metal oxides is intimately linked  to the way the valence electrons occupy the $d$ orbitals~\cite{Tokura2000}. In the cuprates where Cu has a nominal $3d^{9}$ configuration, the hole resides in a single $d_{x^2-y^2}$ orbital on each Cu ion, and much of their physics can be understood from this single-band starting point~\cite{Bednorz1986,Zhang1988,Keimer2015}. Whether high-temperature superconductivity requires such a single-orbital configuration, or can also emerge when holes are distributed across multiple orbitals, has long remained an open question. The bilayer Ruddlesden--Popper nickelates now offer an opportunity to address this question. In \lno, superconductivity appears under pressure with a critical temperature $T_c$ approaching 80~K~\cite{Sun2023}. Rare-earth substitution raises $T_c$ to 82~K in the bulk superconductor La$_2$PrNi$_2$O$_7$~\cite{Wang2024Pr} and to 92~K in \lsno\ and 96~K in the Sm-richer La$_{1.57}$Sm$_{1.43}$Ni$_2$O$_{7-\delta}$~\cite{Li2025Sm}, and compressive strain stabilizes superconductivity at ambient pressure in thin films~\cite{Ko2024,Zhou2025,Wang2024review}.

In these compounds, Ni has a nominal $3d^{7.5}$ configuration and both $e_g$ orbitals contribute to the states at the Fermi level. The $d_{z^2}$ orbitals of the two layers couple through the inner apical oxygen and form bonding and antibonding bands, whereas $d_{x^2-y^2}$ forms a wide in-plane band~\cite{Luo2023,Zhang2023}. Theories differ on which orbital drives the pairing: interlayer superexchange of the $d_{z^2}$ electrons in one class of models~\cite{Yang2023s,Yang2023int,Lu2024} whereas others emphasize the dominant role of the $d_{x^2-y^2}$ states at the Fermi level~\cite{LaBollita2024}. They also start from different orbital fillings. Simple electron counting, with a filled bonding $d_{z^2}$ band, gives 1 $d_{z^2}$ and 0.5 $d_{x^2-y^2}$ electron per Ni, whereas the equal-occupation model of Ref.~\cite{Luo2023} gives about 0.75 electron in each orbital. Electronic correlations can shift this balance: correlated calculations find a strong Ni--O charge transfer toward a dominant Ni $d^8$ configuration~\cite{Shilenko2023,ZhouJ2026}, in which Hund's coupling favors one hole in each $e_g$ orbital. The ratio $\hz/\hx$ of $d_{z^2}$ to $d_{x^2-y^2}$ holes is therefore a direct test of the starting point of these models, ranging from 2/3 for a filled bonding band to about 1 for Hund's-coupled $d^8$.

Measuring this ratio in a bulk crystal is, however, difficult. 
Angle-resolved photoelectron spectroscopy (ARPES) finds a renormalized $d_{z^2}$ flat band 30--50~meV below $E_F$~\cite{Yang2024arpes} or very close to $E_F$~\cite{Li2024cpl}, a difference attributed to oxygen vacancies. But ARPES is surface sensitive, probes quasiparticles near $E_F$ rather than the orbital density integrated over the band, and finds an additional hole-doped $d_{z^2}$ pocket around the M point in superconducting thin films~\cite{Li2025film}. Electron-energy-loss spectroscopy (EELS) finds ligand holes on the inner apical and planar oxygens~\cite{Dong2024}, and RIXS identifies Ni $e_g$ and O $2p$ states at low energy~\cite{Chen2024rixs}. X-ray linear dichroism at the O $K$ edge shows holes in both $e_g$ orbitals in thin films~\cite{Ren2025}, and polarization-dependent x-ray absorption shows a weak orbital polarization in bulk samples~\cite{He2026}, but a quantitative sum-rule analysis~\cite{Wu2013} is complicated by the Ni $L$-edge multiplets and by the overlapping La $M_4$ edge.

Non-resonant inelastic x-ray scattering from an $s$ core level ($s$-NIXS) overcomes these limitations~\cite{Yavas2019}. It uses hard x-rays and is therefore bulk sensitive. At large momentum transfer $\mathbf q$, the dipole-forbidden $3s\to3d$ (Ni $M_1$) transition dominates. Because the $3s$ core hole is spherically symmetric, the dependence of the energy-integrated intensity on the direction $\hat{\mathbf q}$ directly images the angular distribution of the $3d$ holes, free from multiplet effects. The method was demonstrated in NiO~\cite{Yavas2019} and recently applied to perovskite, infinite-layer and single-layer nickelates~\cite{AbarcaMorales2026}. Here we use $s$-NIXS on a bulk \lsno\ crystal to measure $\hz/\hx$. We find a ratio close to 1, well above the value 2/3 of a filled bonding band, which points to Hund's-coupled Ni $d^8$ with ligand holes.

\begin{figure*}[t]
\centering
\includegraphics[width=\textwidth]{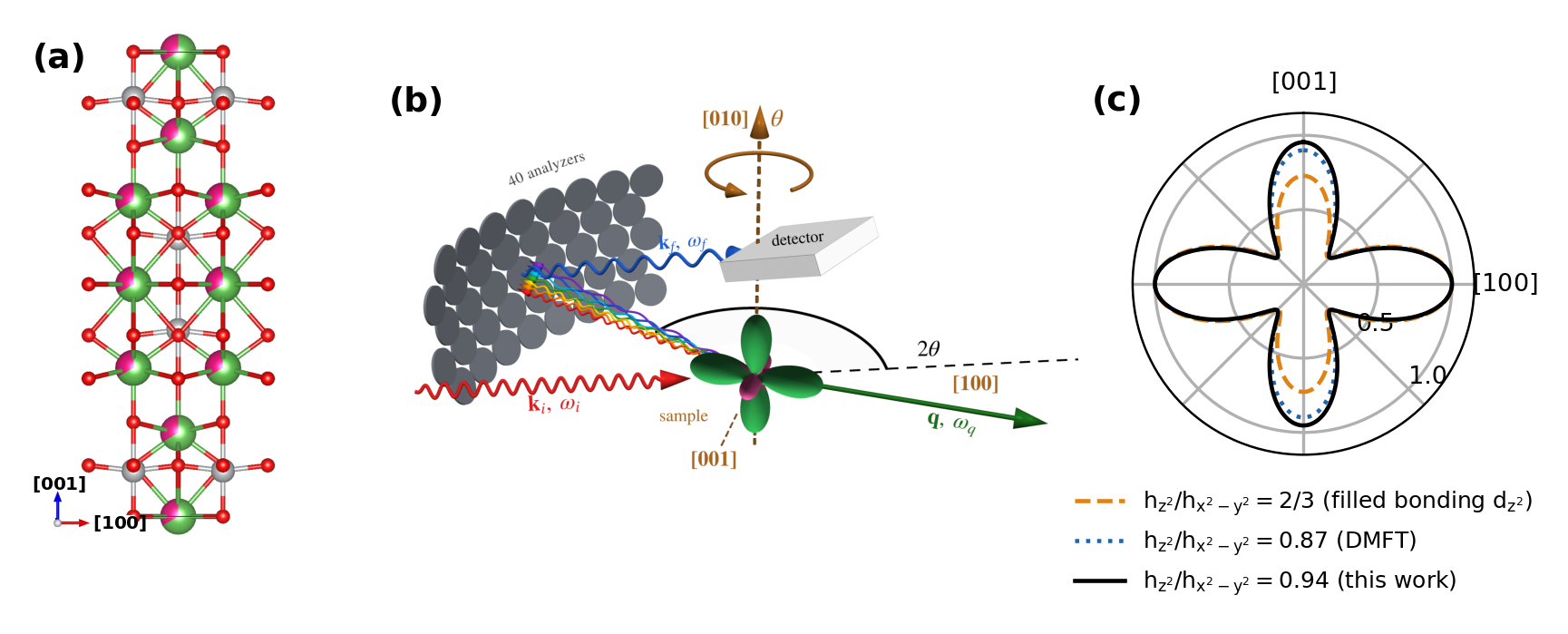}
\caption{\label{fig1}(a) Average tetragonal structure of \lsno\ ($I4/mmm$, no.~139), viewed along $[010]$ with the axes $[100]$ and $[001]\parallel c$ (as in panels b and c); the actual symmetry is monoclinic $P2_1/m$~\cite{Li2025Sm}. Ni gray, O red, La/Sm green/pink, drawn with the average occupancy (La 2/3, Sm 1/3) on all rare-earth sites. (b) Scattering geometry. The incident beam ($\mathbf k_i,\omega_i$) is scattered by the sample; the analyzers select the energy $\omega_f$ of the scattered beam ($\mathbf k_f$) and image it on a two-dimensional detector. The momentum transfer $\mathbf q=\mathbf k_i-\mathbf k_f$ lies in the horizontal plane. The sample rotates about the vertical axis $y=[010]$, so that $\mathbf q$ sweeps the $(x,z)$ plane; $\theta$ denotes the angle between $\mathbf q$ and $z\parallel c$. The shapes drawn at the sample are the angular densities $|Y_{z^2}|^2$ (green), along $z$, and $|Y_{x^2-y^2}|^2$ (pink), along $x$ and $y$. Only the ten analyzers closest to backscattering are used here. (c) Angular dependence of the Ni $M_1$ intensity from Eq.~(\ref{eq1}) for $\hz/\hx=2/3$ (filled bonding $d_{z^2}$ band), 0.87 (DMFT~\cite{Shilenko2023}) and 0.94 (this work), normalized at $\theta=90^\circ$.}
\end{figure*}

The single crystal ($50\times50\times110~\mu$m$^3$) was grown at ambient pressure from a K$_2$CO$_3$ flux, as described in Ref.~\cite{Li2025Sm} which also reports its superconductivity under high pressure. At ambient pressure the structure is monoclinic ($P2_1/m$), with Sm preferentially occupying the rare-earth site between the bilayers~\cite{Li2025Sm,Huang2026}. Prior to the $s$-NIXS measurements, the crystal was oriented by x-ray diffraction within the average tetragonal space group $I4/mmm$ (139), with $a\simeq3.8$~\AA\ and $c\simeq20$~\AA\textcolor{red}{~\cite{SM}}. In the following, $x=[100]$ and $y=[010]$ run along the Ni--O bonds and $z=[001]\parallel c$. The azimuthal orientation is known to within $0.5^\circ$, so that a misalignment reduces the $d_{x^2-y^2}$ contribution according to $\cos^22\varphi$ (with $\varphi$ the deviation angle from perfect alignment) to less than $10^{-3}$.

The measurements were performed at room temperature on the GALAXIES beamline at SOLEIL Synchrotron using the MATRIXS spectrometer equipped with 40 Si(110) spherically bent analyzer crystals ($R=1$~m). The spectrometer operates in a fixed Johann geometry with a Bragg angle of 87$^\circ$. In this geometry, the analyzer energy is fixed at 9.70~keV, while the incident energy is scanned from 9.785 to 9.825~keV across the Ni $M1$ edge energy. The overall energy resolution is 1~eV at $|\mathbf q|=9.65$~\AA$^{-1}$ as estimated from the width of the elastic line. Of the 40 analyzers, the ten analyzers closest to the backscattering direction ($2\vartheta=155^\circ$) were selected, and their signals summed. The resulting solid angle covers $\pm10^\circ$ both within and perpendicular to the scattering plane, corresponding to an angular spread of about $\pm5^\circ$ in $\hat{\mathbf q}$ (where $\hat{\mathbf q}$ is the mean direction over the selected analyzers) since a rotation $\delta$ of the scattered wave vector rotates $\hat{\mathbf q}$ by $\delta/(2\sin\vartheta)$. To obtain the orbital symmetry, the crystal was rotated about the $y$ axis, so that $\hat{\mathbf q}$ sweeps the $(x,z)$ plane at an angle $\theta$ from $z$ [Fig.~\ref{fig1}(b)]. Each spectrum contains the La $N_{4,5}$ and Ni $M_1$ edges. The Sm $N_{4,5}$ edge which falls at approximately 129~eV is not included in our energy loss window.

Figure~\ref{fig2}(a) shows an example of $s$-NIXS spectrum at one angle. The complete data set is provided in the Supplemental Material~\cite{SM}. The spectrum consists of a strong signal from the La $N_{4,5}$ edges between 98 and 107~eV, followed by the Ni $M_1$ edge centered at $110.76\pm0.08$~eV. 
%
%
As the La contribution is isotropic (see Suppl. Mat.), it can can be used to normalize the Ni $M_1$ signal from possible intensity variations arising from changes in the scattering volume upon rotation, as well as variations in absorption and incident-beam intensity. To extract the $s$-NIXS angular dependence, the full spectrum is fitted for each angle using the sum of a quadratic background, a single Gaussian for the Ni peak and the La edge line shape, which was derived from measurements in the Ni-free cuprate La$_{2-x}$Sr$_x$CuO$_4$ and is only scaled in amplitude.

We note that the width of the Ni contribution ($5.57\pm0.22$~eV) is larger than the spectrometer resolution; the implications for the results are discussed below. Figure~\ref{fig2}(b) shows the Ni spectra after subtraction of the La edge and of the background. The integrated Ni intensity exhibits a pronounced angular dependence, with similar values along $z$ and along $x$ and a minimum near $45^\circ$.\\

\begin{figure}[t]
\centering
\includegraphics[width=\columnwidth]{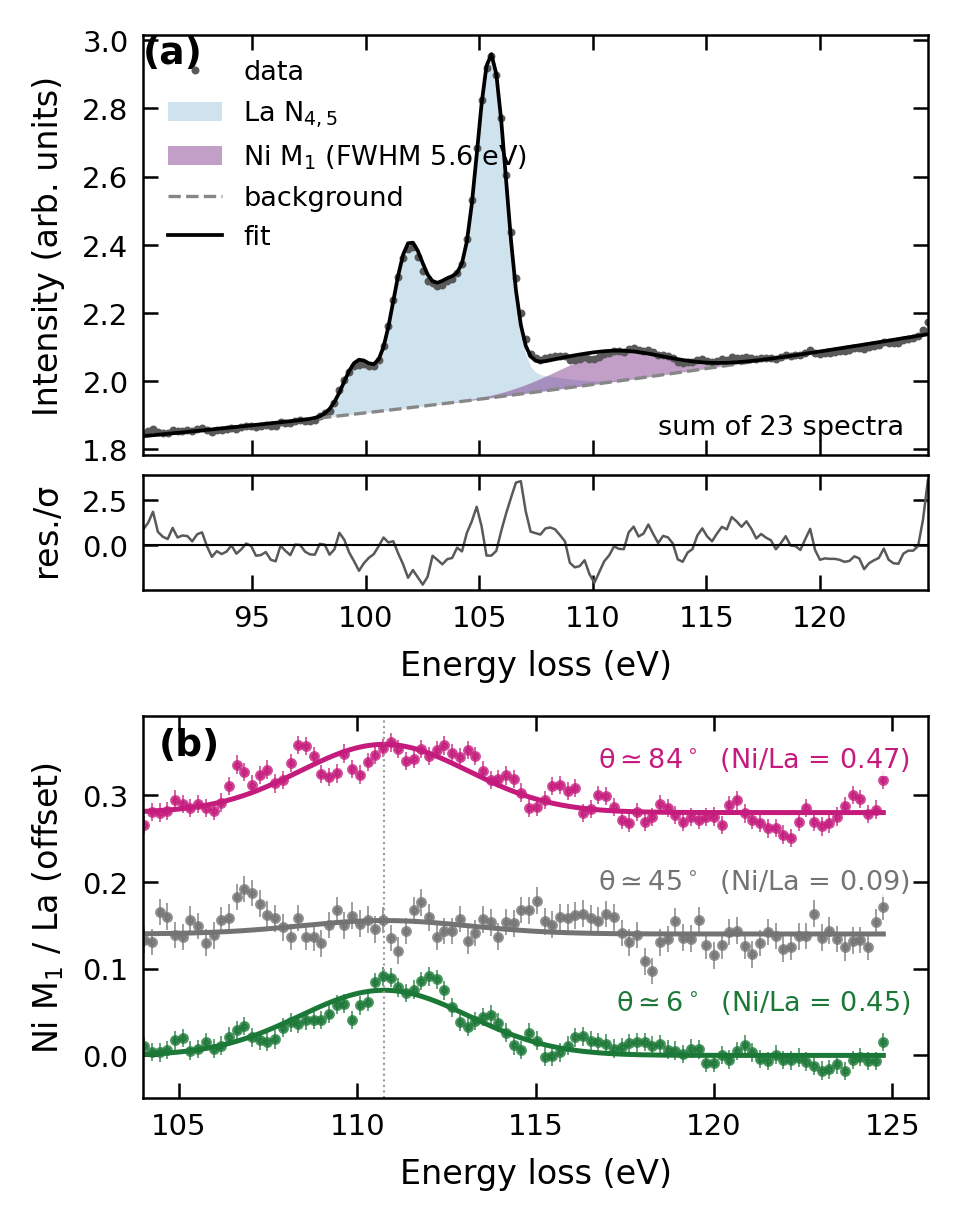}
\caption{\label{fig2}(a) Sum of the energy-loss spectra of \lsno\ (points) with the fit: La $N_{4,5}$ edge (blue, line shape of La$_{2-x}$Sr$_x$CuO$_4$), Ni $M_1$ peak (purple), quadratic background (dashed). Residuals are shown below. (b) Ni $M_1$ peak after subtraction of the La edge and background, for three directions of $\mathbf q$ obtained by symmetry averaging (offset for clarity). Lines: Gaussian of fixed position and width with free area.}
\end{figure}

Considering a hole in an orbital $\psi$,  the $3s\to3d$ intensity~\cite{Haverkort2007,Yavas2019} can be expressed as $|\langle R_{3d}|j_2(qr)|R_{3s}\rangle|^2\,|\psi(\hat{\mathbf q})|^2$. Here $j_2$ is a spherical Bessel function, and $|\psi(\hat{\mathbf q})|^2$ is the angular part of the orbital density. The radial factor is the same for both $e_g$ orbitals and is included into a global scaling factor $I_0$. For $\hat{\mathbf q}$ in the $(x,z)$ plane,

\begin{equation}
I(\theta)=I_0\!\left[\hz\,\frac{5}{16\pi}(3\cos^2\theta-1)^2+\hx\,\frac{15}{16\pi}\sin^4\theta\right],
\label{eq1}
\end{equation}

where $\hz$ and $\hx$ are the numbers of holes in each orbital. Equation~(\ref{eq1}) assumes that the $t_{2g}$ shell is fully occupied. A $t_{2g}$ ($d_{xz}$) hole would add a $\sin^2 2\theta$ term, which reaches a maximum at $45^\circ$. Adding such a term, with a free amplitude $A_t$, to the fit of the data gives $A_t/A_z=-0.03\pm0.06$ relative to the $d_{z^2}$ amplitude: it is compatible with zero, so there is no evidence of $t_{2g}$ holes, and $\hz/\hx$ is unchanged~\cite{SM}. Equation~(\ref{eq1}) also neglects off-diagonal $d_{z^2}$--$d_{x^2-y^2}$ terms. In the low-symmetry structure such terms are allowed. They give a contribution $\propto(3\cos^2\theta-1)\sin^2\theta\cos2\varphi$, which changes sign between twin domains rotated by $90^\circ$. It therefore cancels for equal twin fractions, which is supported by x-ray diffraction on the crystal~\cite{SM}.

\begin{figure}[t]
\centering
\includegraphics[width=\columnwidth]{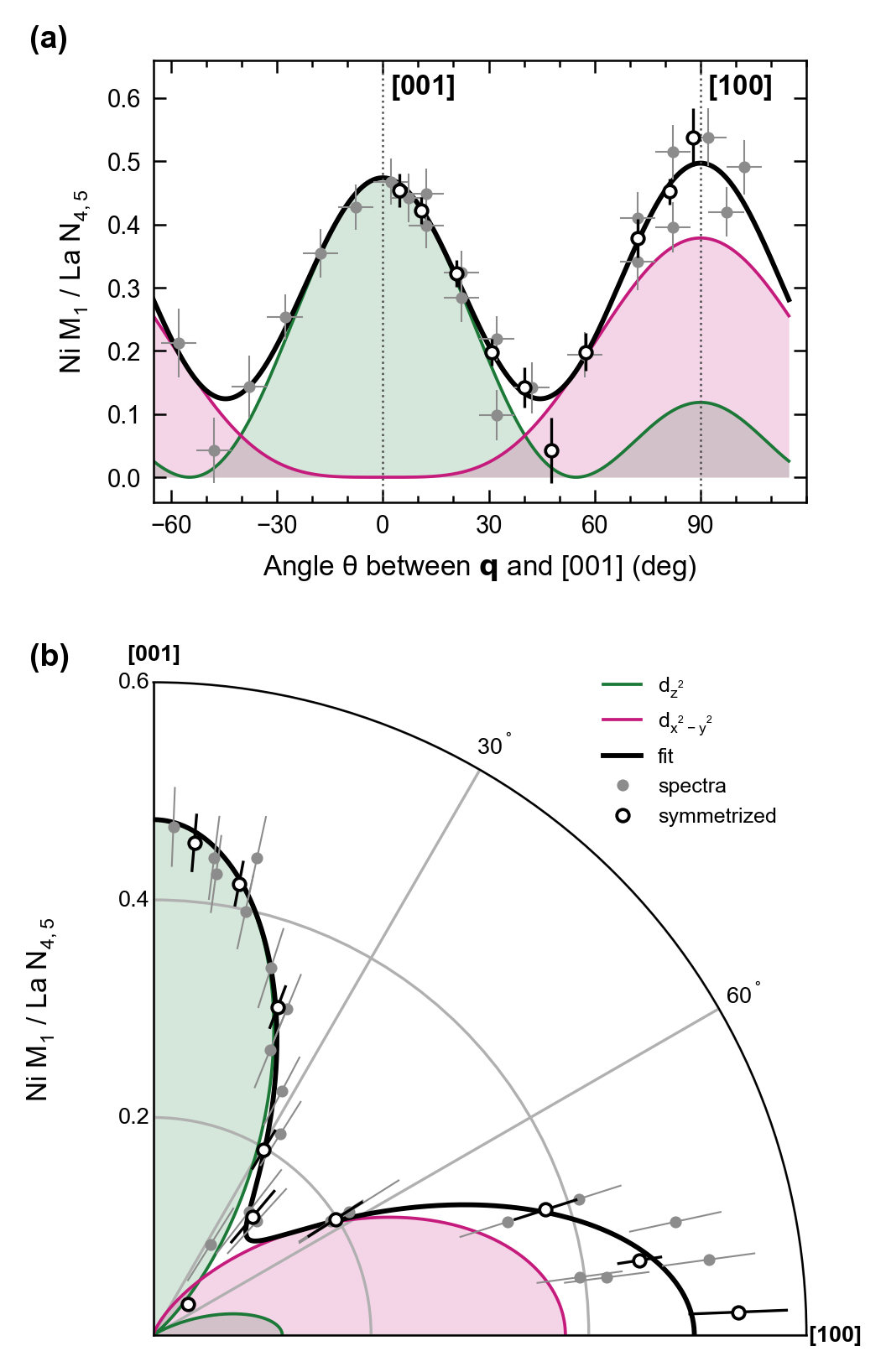}
\caption{\label{fig3}(a) Ni $M_1$/La $N_{4,5}$ intensity of \lsno\ as a function of the angle $\theta$ between $\mathbf q$ and $z=[001]$, over one period of $180^\circ$. Gray: individual spectra (spectra at $\theta$ outside the window are shifted by $180^\circ$; horizontal bars: $\pm5^\circ$ acceptance); open circles: symmetry-averaged spectra. Black line: fit with Eq.~(\ref{eq1}); green and pink: $d_{z^2}$ and $d_{x^2-y^2}$ contributions. (b) The same data in polar coordinates, folded into one quadrant. The fit gives $\hz/\hx=0.94\pm0.06$ with $\chi^2_r=1.09$.}
\end{figure}

Figure~\ref{fig3} shows the measured intensity as a function of the angle $\theta$ between $\mathbf q$ and $z$ [Fig.~\ref{fig3}(a)] and folded into one quadrant by the mirror symmetries $z\to-z$ and $x\to-x$ [Fig.~\ref{fig3}(b)]. The main result is already apparent from the data: After symmetry averaging, the intensities near the $z$ and $x$ directions are nearly equal. The Ni/La intensity ratios are $0.45\pm0.03$ at $\theta\simeq5^\circ$ and $0.45\pm0.02$ at $\theta\simeq81^\circ$. The fitted intensity contrast, $I(z)/I(x)=0.95\pm0.05$, is substantially larger than the value of 0.73 expected for a hole occupancy ratio $\hz/\hx=2/3$. The intensity reaches a minimum near $\theta=45^\circ$.

We fit Eq.~(\ref{eq1}) to the data after averaging over the $\pm5^\circ$ acceptance angle using three free parameters: $A_z=I_0\hz$, $A_x=I_0\hx$ and an angular offset $\theta_c$. We obtain $A_z=1.19\pm0.04$, $A_x=1.27\pm0.06$ and $\theta_c=7.7\pm1.0^\circ$, with a reduced $\chi^2$ of 1.09; the uncertainties are scaled by $\sqrt{\chi^2_r}$. The offset is close to the $6.6^\circ$ expected from the orientation of the crystal determined by x-ray diffraction. Including the correlation between $A_z$ and $A_x$, we find

\begin{equation}
\hz/\hx=0.94\pm0.06\,(\text{stat})\,^{+0.08}_{-0.01}\,(\text{syst}).
\label{eq2}
\end{equation}

This corresponds to a fraction $48.4\pm1.6\%$ of the $e_g$ holes in $d_{z^2}$. A fit of the symmetry-averaged data gives $0.93\pm0.06$. The systematic range comes from varying one analysis choice at a time~\cite{SM}: the selection thresholds, the angular acceptance, the angular offset, the background and the Ni line shape (Fig.~S3 of Ref.~\cite{SM}). The largest change, $+0.08$, comes from a cubic background. The ratio also depends on the assumed width of the Ni peak, by $-0.11$ per eV (1.02, 0.91 and 0.83 for imposed widths of 5, 6 and 7~eV). The width is determined by the data, and its statistical uncertainty, $\pm0.22$~eV, changes the ratio by $\pm0.02$~\cite{SM}.

If all $e_g$ holes were localized on Ni and the bonding $d_{z^2}$ band were fully occupied, each Ni would have one $d_{z^2}$ hole and 1.5 $d_{x^2-y^2}$ holes, yielding $\hz/\hx=2/3$. Our measured value exceeds this by 4.4 statistical standard deviations, and constraining the ratio to 2/3 deteriorates the fit $\chi^2_r$ value from 1.09 to 2.53. All alternative analyses yield values above 2/3, the lowest being $0.93\pm0.06$. The same conclusion holds when the Ni peak width is fixed to values between 4 and 7~eV; the lowest ratio in this range is $0.83\pm0.05$, obtained for a width of 7~eV. Our result is consistent with the DMFT prediction of $\hz/\hx\simeq0.87$~\cite{Shilenko2023}, which is obtained from Wannier occupations of 0.60 ($d_{z^2}$) and 0.54 ($d_{x^2-y^2}$) per spin orbital. Our accuracy does not distinguish this value from 1.

A ratio close to 1 can be understood within an atomic picture. In presence of strong Hund's coupling, a high-spin ($S=1$) Ni $d^8$ ion has one hole in each $e_g$ orbital, yielding $\hz/\hx=1$. In contrast, a low-spin $d^7$ ion with its $e_g$ electron in $d_{z^2}$ yields a ratio of 0.5 while a filled bonding band yields a ratio of 2/3. Our result therefore suggests a Ni configuration close to $d^8$ with the remaining holes residing on the oxygen ligands. This interpretation is consistent with the Ni $3d$ count of approximately 8.24 obtained in DMFT calculations~\cite{Shilenko2023} indicating a dominant contribution from the $d^8$ configuration~\cite{ZhouJ2026}. DFT$+U$ calculations of \lno\ likewise find a Ni $3d$ charge close to $d^8$ on all Ni sites, the $d^7$ character of the low-spin sites appearing only in the spin density~\cite{LaBollita2024}. In this picture, Hund's coupling rather than the bonding--antibonding splitting of $d_{z^2}$ governs the $e_g$ occupations. This picture is consistent with the holes found in both $e_g$ orbitals by linear dichroism at the O $K$ edge in films~\cite{Ren2025} and with the weak orbital polarization observed in bulk samples, also interpreted in terms of Hund's-coupled $d^8$~\cite{He2026}. Our finding adds a quantitative estimate in a bulk crystal without the difficulties associated with multiplet effects and proximity of the La $M_4$ edge.

The DMFT calculation also points to the origin of this deviation. The total Ni $3d$ count is approximately 8.24 compared to the nominal value of 7.5~\cite{Shilenko2023}. Compared with the ionic limit, most of the additional electrons go into $d_{x^2-y^2}$ (0.58 per Ni, against 0.2 in $d_{z^2}$). These extra Ni $3d$ electrons are transferred from the neighboring oxygen, leading to ligand holes with mainly in-plane $x^2-y^2$ symmetry, which is reminiscent of Zhang--Rice singlet states in cuprates~\cite{Zhang1988}. This picture is consistent with our measured orbital hole-occupancy ratio. Two caveats should, however, be considered. First, the calculations are performed in the high-pressure phase; to our knowledge, no equivalent calculation has been reported at ambient pressure. Second, electron-energy-loss spectroscopy finds ligand holes with comparable densities on the planar and on the inner apical oxygens~\cite{Dong2024}; the latter hybridize with $d_{z^2}$ and act in the opposite direction. This is an intrinsic effect, already included in our measurement. Finally, oxygen vacancies, located mainly on the inner apical site~\cite{Dong2024}, donate electrons, which fill the ligand holes on the planar and inner apical oxygens almost equally~\cite{Dong2024}. A slight oxygen deficiency is therefore not expected to bias the ratio in a definite direction.

The quantity measured here is the \emph{hole} ratio $\hz/\hx$; the orbital electron occupations, $n_i=2-h_i$, also depend on the total number of $e_g$ holes $h=\hz+\hx$, which is not measured. Between the ionic value $h=2.5$ and the DMFT value $h\simeq1.72$, our hole ratio of 0.94 yields an electron-occupation ratio $n_{x^2-y^2}/n_{z^2}\simeq0.90$--0.95, and no lower than 0.73 for the lowest hole ratio found when the width of the Ni peak is varied (0.83, for 7~eV). By comparison, a hole ratio of 2/3 (a fully occupied bonding $d_{z^2}$ band) gives $n_{x^2-y^2}/n_{z^2}=0.5$ for $h=2.5$, and the equal-occupation picture of Ref.~\cite{Luo2023} gives $n_{x^2-y^2}/n_{z^2}=1$, i.e.\ $\hz/\hx=1$, independently of $h$. Our result is close to the latter estimate, lending strong support to scenarios in which both orbitals take part in the pairing~\cite{Yang2023int,Lu2024}.

In summary, we have probed the Ni $e_g$ holes distribution in the bilayer nickelate \lsno\ using $s$-NIXS. We obtain a hole occupation ratio $\hz/\hx=0.94\pm0.06\,^{+0.08}_{-0.01}$, significantly above the value of 2/3 expected for a filled-bonding-band and consistent with charge-transfer DMFT calculations. Our results suggest that theories of superconductivity in these materials should start from nearly equal $e_g$ occupations at ambient pressure. A cluster or multiplet calculation of the Ni $M_1$ final states and a DFT+DMFT calculation of \lsno\ at ambient pressure would allow a more direct comparison with our data, since the available DMFT occupations were obtained for the high-pressure phase~\cite{Shilenko2023}.

$s$-NIXS has several limitations still. First, it probes the Ni-projected hole density averaged over the fluctuating $d^n$ configurations while providing no information about band dispersion. Second, the method remains relative as determining the absolute number of holes $h$ would require a reference compound to be measured using the same setup. Finally, our measurements were performed at room temperature, above the density-wave transition near 150~K reported in \lno~\cite{Zhao2024}.
Nevetheless, our approach provides a model-free description of the orbital physics in bulk correlated systems which applies well to nickelates. In these systems, the method could be extended to the full rare-earth series where chemical pressure is predicted to alter the $e_g$ occupations~\cite{Zhang2023trends,Pan2024,Yan2026}, or different sample environments such as low temperature to cross the density-wave transition or under applied pressure in a diamond anvil cell.

\begin{acknowledgments}
We acknowledge SOLEIL for provision of beamtimes at GALAXIES beamline (Proposals 20210421, 20221287, 20231963, 20240332, 20250647 and 99260232). This work was supported by the Paris Ile-de-France Region within the DIM MaTerRE (project DAC-VX) and DIM-MAP (project MATRIXS-4H) programs. Junjie Zhang was supported by the Natural Science Foundation of Shandong Province, China (Grant No. ZR2026ZD32) and the National Natural Science Foundation of China (Grants Nos. 12374457 and 12074219).
\end{acknowledgments}

\emph{Data availability.} The data that support the findings of this article are available from the authors upon reasonable request.

\bibliography{Biblio}

\end{document}


\title{Supplemental Material for ``Direct imaging of the Ni $e_g$ hole orbitals in the bilayer nickelate \lsno''}

\author{Ethan Saïman}
\affiliation{Université Paris-Saclay, CNRS, Laboratoire de Physique des Solides, 91405 Orsay, France}
\affiliation{Synchrotron SOLEIL, L'Orme des Merisiers, Saint Aubin BP 48, 91192 Gif-sur-Yvette, France}

\author{Feiyu Li}
\affiliation{State Key Laboratory of Crystal Materials, Institute of Crystal Materials, Shandong University, Jinan, Shandong 250100, China}
\affiliation{College of Materials Science and Engineering, Qingdao University of Science \& Technology, Qingdao, Shandong 266042, China}

\author{Junjie Zhang}
\affiliation{State Key Laboratory of Crystal Materials, Institute of Crystal Materials, Shandong University, Jinan, Shandong 250100, China}

\author{Jean-Pascal Rueff}
\affiliation{Synchrotron SOLEIL, L'Orme des Merisiers, Saint Aubin BP 48, 91192 Gif-sur-Yvette, France}
\affiliation{Sorbonne Université, CNRS, Laboratoire de Chimie Physique – Matière et Rayonnement, LCPMR, F-75005 Paris, France}

\author{Victor Balédent}
\email{victor.baledent@universite-paris-saclay.fr}
\affiliation{Université Paris-Saclay, CNRS, Laboratoire de Physique des Solides, 91405 Orsay, France}
\affiliation{Institut universitaire de France (IUF)}

\maketitle

\section{Sample}

The \lsno\ single crystal ($50\times50\times110~\mu$m$^3$) was grown at ambient pressure from a K$_2$CO$_3$ flux at Shandong University, following Ref.~\cite{Li2025Sm}. Its structure is monoclinic ($P2_1/m$), with Sm preferentially on the rare-earth site between the bilayers~\cite{Li2025Sm}. It was oriented by x-ray diffraction at the Laboratoire de Physique des Solides, with an accuracy better than $0.5^\circ$ on the azimuthal angle; the corresponding reduction of the $d_{x^2-y^2}$ term, $\cos^22\varphi$, is below $10^{-3}$.
We use pseudo-tetragonal axes, with $x=[100]$ and $y=[010]$ along the Ni--O bonds and $z=[001]\parallel c$.

\section{X-ray diffraction}\label{sec:xrd}
The crystal used for the $s$-NIXS measurements was characterized on the high-pressure diffraction setup developed at the Laboratoire de Physique des Solides (Orsay, France). The diffractometer uses a microfocus Mo x-ray tube (Incoatec GmbH) with Montel multilayer focusing optics, which delivers a monochromatic beam ($\lambda=0.711$~\AA) with a $100~\mu$m spot size and a divergence of 0.5~mrad at the sample. The diffracted intensity is recorded on a large-area MAR345 image-plate detector (MarXperts GmbH). Data were acquired in steps of $1^\circ$ over a rotation range of $340^\circ$, with an exposure of 300~s per frame. The images were indexed with CrysAlisPro~\cite{CrysAlisPRO2022}: 1017 of the 1252 measured Bragg reflections are indexed with a tetragonal cell of $5.4\times5.4\times20.16$~\AA$^3$. This cell is the $\sqrt2\times\sqrt2$ cell of the $I4/mmm$ cell used in the main text ($a=5.4/\sqrt2\simeq3.8$~\AA, $c\simeq20.2$~\AA).

The tetragonal average structure shows that the crystal is not a single monoclinic domain, the monoclinic distortion of the $P2_1/m$ structure being averaged over domains rotated by $90^\circ$ relative to each other, with comparable populations. This is the condition for which the off-diagonal $d_{z^2}$--$d_{x^2-y^2}$ terms cancel.

\section{Data acquisition and reduction}

The measurements were performed at room temperature on the GALAXIES beamline of SOLEIL. The spectrometer has 40 spherically bent Si(110) analyzers ($R=1$~m), imaged on a two-dimensional detector, which select a fixed final energy of 9.70~keV. The incident energy is scanned from 9.785 to 9.825~keV with 161 points and 30~s per point (10~s per point for the scans at nominal angles $-100^\circ$ to $-60^\circ$); the energy resolution is 1~eV. For this work we use the ten analyzers closest to backscattering, centered at $2\vartheta=155^\circ$ ($|\mathbf q|=9.65$~\AA$^{-1}$), which give the best statistics.

For each angle, an elastic scan and an energy-loss scan were recorded. The data set contains 36 energy-loss scans at 26 nominal angles; repeated scans at the same angle are treated as independent points. For each scan, the procedure is as follows:
\begin{enumerate}
\item Hot and dead pixels are replaced by the median of their neighbors.
\item One region of interest is defined for each analyzer and is re-centered on the elastic image of the same scan.
\item The energy scale of each analyzer is calibrated on the elastic line of the same scan (or, if absent, on the first elastic scan at that angle).
\item The ten analyzer spectra are interpolated on a common energy axis and summed.
\end{enumerate}
No normalization by a beam monitor is applied. The normalization is provided by the La $N_{4,5}$ area (Sec.~\ref{sec:lineshape}). This area is an internal reference measured simultaneously with the Ni signal, on the same detector pixels and from the same scattering volume.

\section{Line-shape model}\label{sec:lineshape}

\subsection{La $N_{4,5}$ edge}
The $4d\to4f$ excitation of La$^{3+}$ connects a full $4d$ shell to an empty $4f$ shell. It is not purely dipolar: at large $q$ the operator $e^{i\mathbf q\cdot\mathbf r}$ contains all multipoles of rank $k=1$, 3 and 5 allowed between $l=2$ and $l=3$, with $q$-dependent weights (radial integrals of $j_k(qr)$). Because the initial state is a closed shell ($^1S_0$, $J=0$), a multipole of rank $k$ can only reach final states with $J_f=k$, so that different ranks do not interfere, and for each of them the sum over $M_f$ of $|Y_{kM_f}(\hat{\mathbf q})|^2$ is independent of $\hat{\mathbf q}$. Non-dipolar contributions therefore modify the shape of the edge, and not its isotropy. As the weights depend on $|\mathbf q|$, the line shape must be measured at the same $|\mathbf q|$ as the nickelate, which is the case for La$_{2-x}$Sr$_x$CuO$_4$ (Sec.~\ref{sec:lacheck}). The intensity of the whole edge is therefore independent of the direction of $\mathbf q$ at every energy loss, apart from the crystal-field splitting of the $4f$ final states (tens of meV), which is negligible at the 1~eV resolution. The edge (98--107~eV) lies entirely within the fit window (88--126~eV) and is fitted as a whole. Its shape is described by five Gaussians (four peaks and a shoulder at 107~eV) determined on La$_{2-x}$Sr$_x$CuO$_4$, a La compound without Ni (Sec.~\ref{sec:lacheck}, Table~\ref{tab:la}). In the fit of each spectrum, this shape is shifted rigidly by 0.06~eV (determined on the sum of the retained spectra; the allowed range is $\pm0.5$~eV), and only a global scale $S$ is free. The quantity analyzed in the main text is the Ni $M_1$ area divided by the area of the main La Gaussian times $S$. The Sm $N_{4,5}$ edge sets in at about 126~eV, with a maximum at 129~eV, at the upper end of the window.

\begin{table}[h]
\caption{\label{tab:la}La $N_{4,5}$ line shape determined on La$_{2-x}$Sr$_x$CuO$_4$ (Gaussians; areas relative to the main peak; positions before the 0.06~eV shift).}
\begin{ruledtabular}
\begin{tabular}{cccc}
Peak & Position (eV) & FWHM (eV) & Relative area \\
\hline
0 & 105.46 & 1.56 & 1 \\
1 & 101.84 & 1.75 & 0.550 \\
2 & 99.60 & 1.52 & 0.155 \\
3 & 103.64 & 1.73 & 0.364 \\
4 (shoulder) & 107.12 & 3.85 & 0.122 \\
\end{tabular}
\end{ruledtabular}
\end{table}

\subsection{Check of the La $N_{4,5}$ line shape on a Ni-free compound}\label{sec:lacheck}
\begin{figure}[h]
\includegraphics[width=0.6\textwidth]{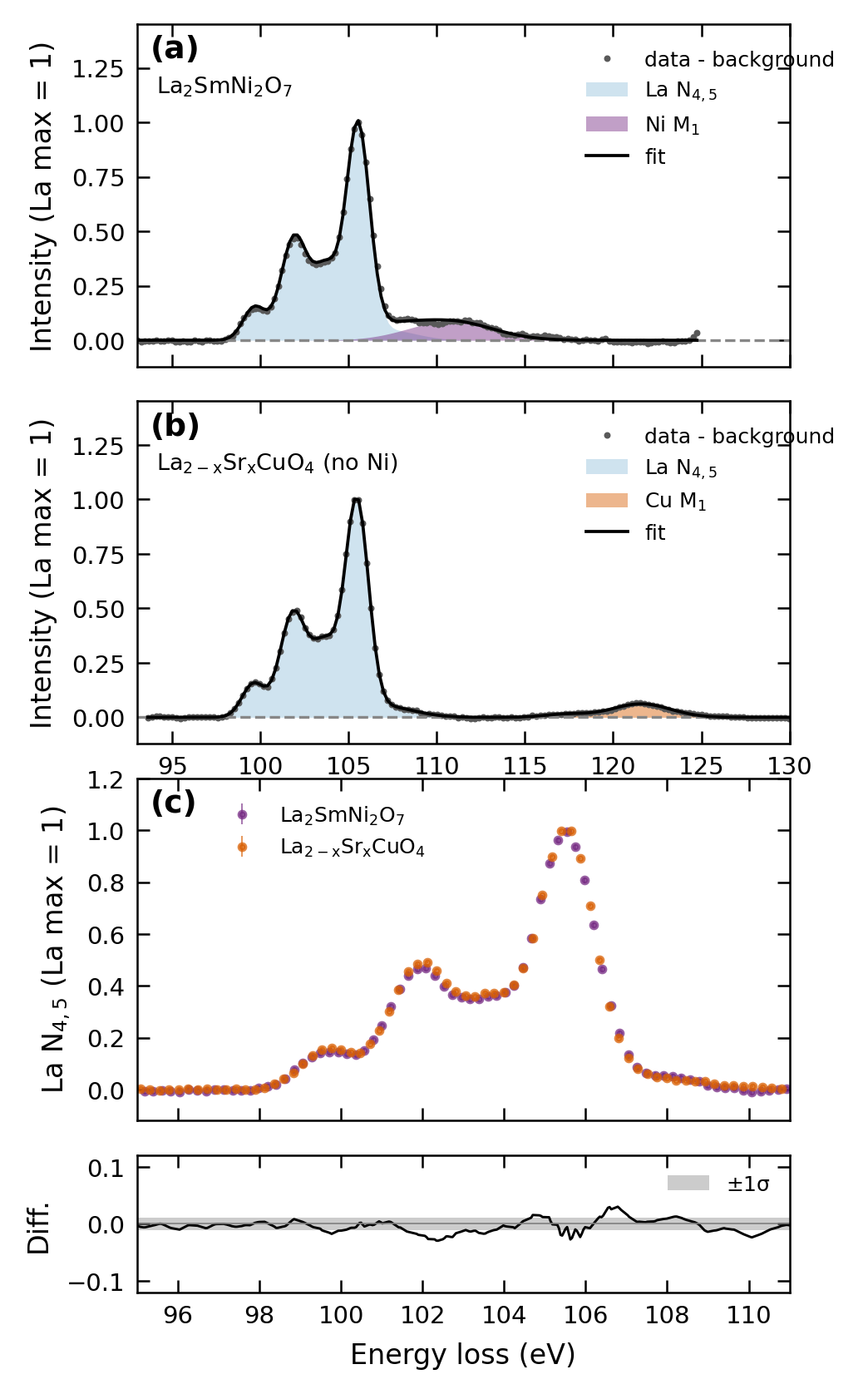}
\caption{\label{fig:lacheck}Check of the La $N_{4,5}$ line shape. (a) \lsno: data minus background (points), La $N_{4,5}$ edge (blue) and Ni $M_1$ peak (purple); the La maximum is set to 1. (b) La$_{2-x}$Sr$_x$CuO$_4$, a La compound without Ni, in the same units: La $N_{4,5}$ edge (blue) and Cu $M_1$ peak (orange). (c) La $N_{4,5}$ edges of the two compounds (data minus background and $M_1$ peak), normalized to their maximum; the energy of the cuprate is shifted by $0.06$~eV. Bottom: difference between the two (nickelate minus cuprate); the gray band is the noise ($\pm1\sigma$).}
\end{figure}

The La edge is the reference for the whole analysis, so its shape must not depend on the sample. We determined it on a compound that contains La but no Ni, the cuprate La$_{2-x}$Sr$_x$CuO$_4$ ($x=0.08$), measured at 300~K in the same setup, with the same analyzers and the same scattering angle ($2\vartheta=155^\circ$). Its spectrum has the same La $N_{4,5}$ edge and, at higher energy, the Cu $M_1$ peak at 121.6~eV (Fig.~\ref{fig:lacheck}b). The spectrum of this compound is the sum of all its measurements, fitted with La $N_{4,5}$ (five Gaussians, Table~\ref{tab:la}), a parabolic background and the $M_1$ peak ($\chi^2_r=1.67$). In Fig.~\ref{fig:lacheck}c we superimpose the La $N_{4,5}$ edge of the two compounds, i.e.\ the data minus the fitted background and $M_1$ peak, each normalized to the maximum of the edge. The two edges coincide in position and in the relative weights of the three structures at 99.6, 102 and 105.5~eV. After the rigid energy shift of $0.06$~eV used in the fits, which is within the calibration uncertainty, the rms difference between 97 and 108~eV is 1.3\% of the maximum, and the largest deviation is 3\% (Fig.~\ref{fig:lacheck}c, bottom), i.e.\ 1.2 times the noise. These deviations are of the order of the noise and of the uncertainty on the subtraction of the $M_1$ tail. The La edge of \lsno\ therefore has the shape of an isolated La$^{3+}$ $4d\to4f$ edge, and the shape of Table~\ref{tab:la}, which is used for the nickelate with only a shift and an amplitude, does not contain a Ni contribution. We use this comparison only for the La line shape, not for the $M_1$ peaks of the two compounds.

\subsection{Background and Ni $M_1$ peak}
The background is a quadratic polynomial over the full window; linear and cubic backgrounds are tested in Table~\ref{tab:syst}. The Ni $M_1$ peak is a single Gaussian of non-negative amplitude. Its position, $110.76\pm0.08$~eV, and width, FWHM $=5.57\pm0.22$~eV, are fixed on the sum of the retained spectra at nominal angles $-20^\circ$ to $30^\circ$, i.e.\ with $\hat{\mathbf q}$ within $\simeq25^\circ$ of $z$, where the peak is strongest. For each spectrum, only the Ni area $A\geq0$, the La scale $S$ and the background are free. The noise of each spectrum is estimated from point-to-point differences. The uncertainty on $A/S$ is obtained from the covariance matrix, including the $A$--$S$ correlation. The width exceeds the 1~eV resolution; its effect on the hole ratio is discussed in Sec.~\ref{sec:syst}.

\section{Selection of the spectra}\label{sec:selection}
Of the 36 spectra, 13 are discarded. Ten of them belong to the block of nominal angles $-100^\circ$ to $-60^\circ$, where the counting time was 10~s per point instead of 30~s, i.e.\ three times shorter. Six of these have an uncertainty on Ni/La above 1.2 times the median (0.045, for a median of 0.037). Three scans at $-65^\circ$ and $-60^\circ$, whose uncertainty is below that threshold, are excluded with the rest of the block. The tenth ($-70^\circ$\#2) is rejected as an outlier (see below). Two spectra at $-20^\circ$ have an uncertainty of 0.070 and 0.091, above the threshold. After a first robust fit of the orbital model, spectra whose residual exceeds 2.5 robust standard deviations (median absolute deviation) are rejected iteratively; this removes $-70^\circ$\#2 and one spectrum at $160^\circ$. The spectra at $-50^\circ$, $-40^\circ$ and $-30^\circ$ have an uncertainty of 0.046--0.051, slightly above the threshold; they are retained because they lie near the minimum of the angular dependence (without them, $\hz/\hx=0.932\pm0.062$, Table~\ref{tab:syst}). The remaining 23 spectra, at 19 angles, are used. The effect of the thresholds on the hole ratio is given in Table~\ref{tab:syst}.

\section{Orbital fit}
\begin{figure}[h]
\includegraphics[width=0.85\textwidth]{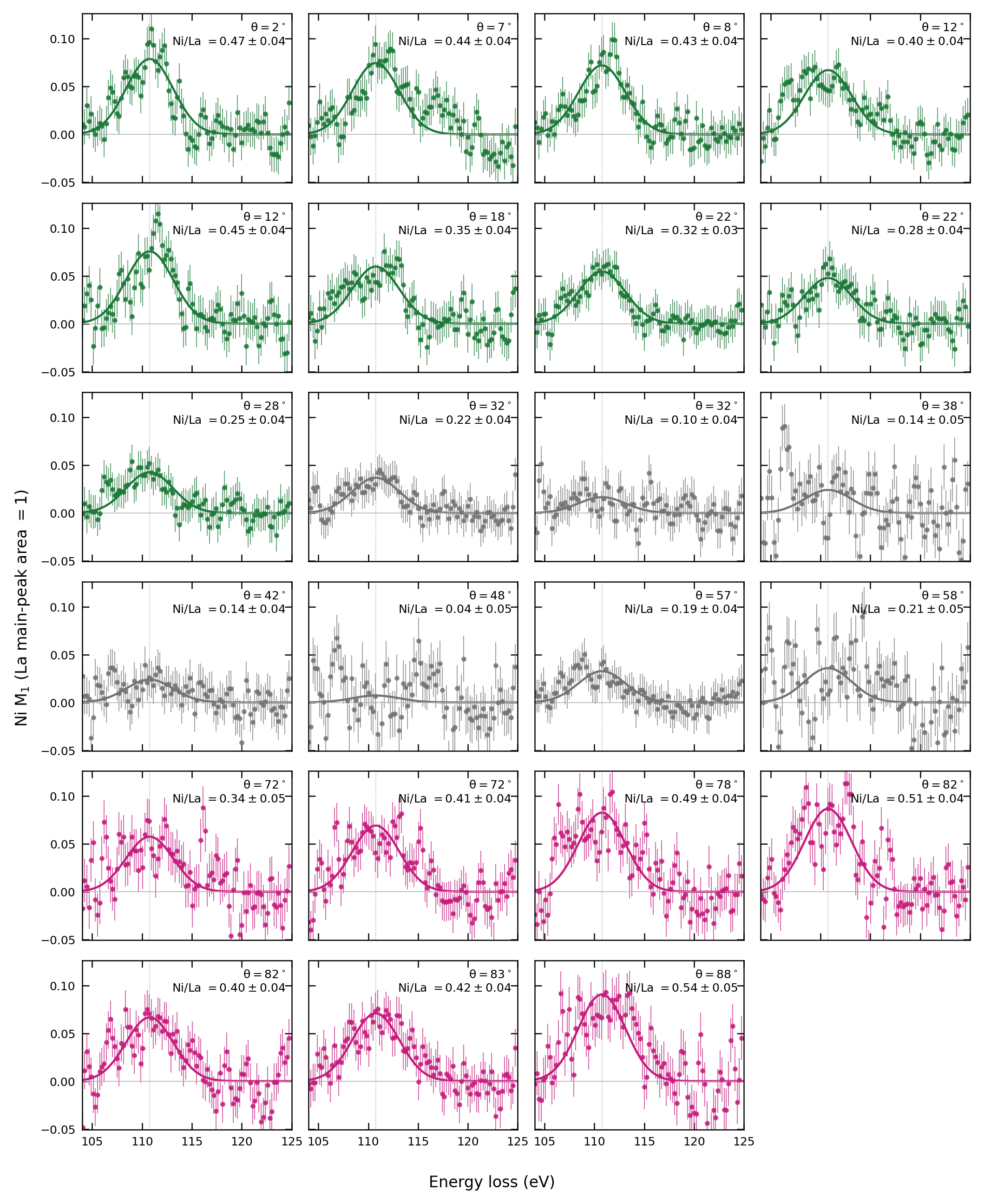}
\caption{\label{fig:allspectra}Ni $M_1$ peak of the 23 retained spectra after subtraction of the La $N_{4,5}$ edge and of the background (points with statistical error bars), with the fitted Gaussian (line; fixed position and width, free area). Each panel is normalized to the La scale of the spectrum (area of the main La Gaussian $=1$). The spectra are ordered by the angle $\theta$ between $\mathbf q$ and $z$ (folded into $[0,90^\circ]$), indicated with the Ni/La ratio in each panel. Green: $\theta<30^\circ$; gray: $30^\circ<\theta<60^\circ$; pink: $\theta>60^\circ$. The dotted line marks the position of the peak, $110.76$~eV.}
\end{figure}

\paragraph*{Individual spectra.}Figure~\ref{fig:allspectra} shows the Ni $M_1$ peak of each of the 23 retained spectra, after subtraction of the fitted La $N_{4,5}$ edge and background, in the same way as Fig.~2(b) of the main text. The spectra are ordered by the angle $\theta$ between $\mathbf q$ and $z$, folded into $[0,90^\circ]$. Each spectrum is normalized to the La scale of its own fit (area of the main La Gaussian equal to 1), so that the area of the Gaussian is the Ni/La ratio listed in each panel. The Gaussian has the position and width fixed on the sum of the spectra (Sec.~\ref{sec:lineshape}), and only its area is free. The ratio is largest for $\mathbf q$ close to $z$ ($\theta\lesssim20^\circ$) and to the $xy$ plane ($\theta\gtrsim70^\circ$), and smallest around $\theta=40$--$50^\circ$, as expected from Eq.~(1) of the main text. The spectra differ in statistical quality, because the signal depends on the direction of $\mathbf q$. Some spectra show structures that the Gaussian does not describe, such as a small dip between 115 and 120~eV or a rise above 123~eV. They come from the imperfect subtraction of the tail of the La edge and of the background; they are the origin of the dependence of the ratio on the width of the Ni peak discussed in Sec.~\ref{sec:syst}.

The Ni/La ratio is fitted with
\begin{equation}
R(\theta)=A_z\,\langle|Y_{z^2}(\theta-\theta_c)|^2\rangle+A_x\,\langle|Y_{x^2-y^2}(\theta-\theta_c)|^2\rangle,
\end{equation}
with $|Y_{z^2}|^2=\frac{5}{16\pi}(3\cos^2\theta-1)^2$ and $|Y_{x^2-y^2}|^2=\frac{15}{16\pi}\sin^4\theta$ in the $(x,z)$ plane. Brackets denote the average over the in-plane acceptance, $\pm5^\circ$ (21 points). The offset $\theta_c$ is free. Weighted least squares give $A_z=1.19\pm0.04$, $A_x=1.27\pm0.06$ and $\theta_c=7.7\pm1.0^\circ$ for 23 points, with $\chi^2_r=1.09$. Uncertainties are scaled by $\sqrt{\chi^2_r}$. Then $\hz/\hx=A_z/A_x=0.939\pm0.061$, including the covariance of $A_z$ and $A_x$ (correlation $-0.25$), and the $d_{z^2}$ fraction is $A_z/(A_z+A_x)=48.4\pm1.6\%$. The ratio differs from 2/3 by 4.4 standard deviations; accordingly, fixing $\hz/\hx=2/3$, with one free amplitude and $\theta_c$ free, raises $\chi^2$ from 21.7 to 53.0 ($\chi^2_r$ from 1.09 to 2.53).

\paragraph*{Symmetry averaging.} The intensity is invariant under $z\to-z$ and $x\to-x$. The retained spectra are folded into $0\le\theta\le90^\circ$ and grouped within $5^\circ$. The spectra of each group are summed and refitted. The ten groups give $\hz/\hx=0.935\pm0.056$ ($\chi^2_r=0.56$). Near $z$ ($\theta\simeq5^\circ$, two spectra) and near $x$ ($\theta\simeq81^\circ$, four spectra), the Ni/La ratios are $0.454\pm0.027$ and $0.452\pm0.021$. The contrast $I(z)/I(x)=4\hz/(\hz+3\hx)$ of the fit is $0.95\pm0.05$; the value expected for $\hz/\hx=2/3$ is 0.73. Figure~\ref{fig:polar360} shows the data of Fig.~3 of the main text without folding.

\section{Angular acceptance}
A rotation $\delta$ of $\mathbf k_f$ rotates $\mathbf q=\mathbf k_i-\mathbf k_f$ by $\delta\,k_f/q=\delta/(2\sin\vartheta)\simeq0.51\,\delta$ at $2\vartheta=155^\circ$. The $\pm10^\circ$ spread of the analyzers thus corresponds to about $\pm5^\circ$ on $\hat{\mathbf q}$, in and out of the scattering plane. The fit includes the in-plane spread. The out-of-plane spread gives $\hat{\mathbf q}$ a small $y$ component and multiplies the $d_{x^2-y^2}$ term by $\cos^22\varphi\simeq1-4\langle\varphi^2\rangle$. Averaging the model over the two-dimensional acceptance ($\pm5^\circ$ in and out of the plane) leaves the hole ratio unchanged ($0.938\pm0.061$) and gives $\chi^2_r=1.08$.

\section{Assumptions of the model}\label{sec:assump}
\begin{itemize}
\item \emph{Full $t_{2g}$ shell.} A $t_{2g}$ hole ($d_{xz}$) would add a contribution $\propto\sin^22\theta$, peaked at $45^\circ$. Adding a free term $A_t\frac{15}{16\pi}\sin^22\theta$ to the fit gives $A_t=-0.03\pm0.07$ ($A_t/A_z=-0.03\pm0.06$), compatible with zero, and $\chi^2_r=1.13$, with $\hz/\hx=0.938\pm0.062$, unchanged.
\item \emph{No $d_{z^2}$--$d_{x^2-y^2}$ coherence.} In the monoclinic structure, a hole orbital $\alpha d_{z^2}+\beta d_{x^2-y^2}$ adds a term $2\alpha\beta\,\frac{5\sqrt3}{16\pi}(3\cos^2\theta-1)\sin^2\theta\cos2\varphi$. It changes sign between twin domains rotated by $90^\circ$ ($\varphi\to\varphi+90^\circ$) and cancels for equal twin fractions.
The x-ray diffraction of the crystal gives a tetragonal average cell, which indicates domains rotated by $90^\circ$ in comparable proportions.

\item \emph{Equal radial factors.} The $3s\to3d$ radial integral $\langle R_{3d}|j_2(qr)|R_{3s}\rangle$ is taken to be the same for both $e_g$ orbitals.
\end{itemize}

\section{Systematic uncertainties}\label{sec:syst}
\begin{figure}[h]
\includegraphics[width=0.45\textwidth]{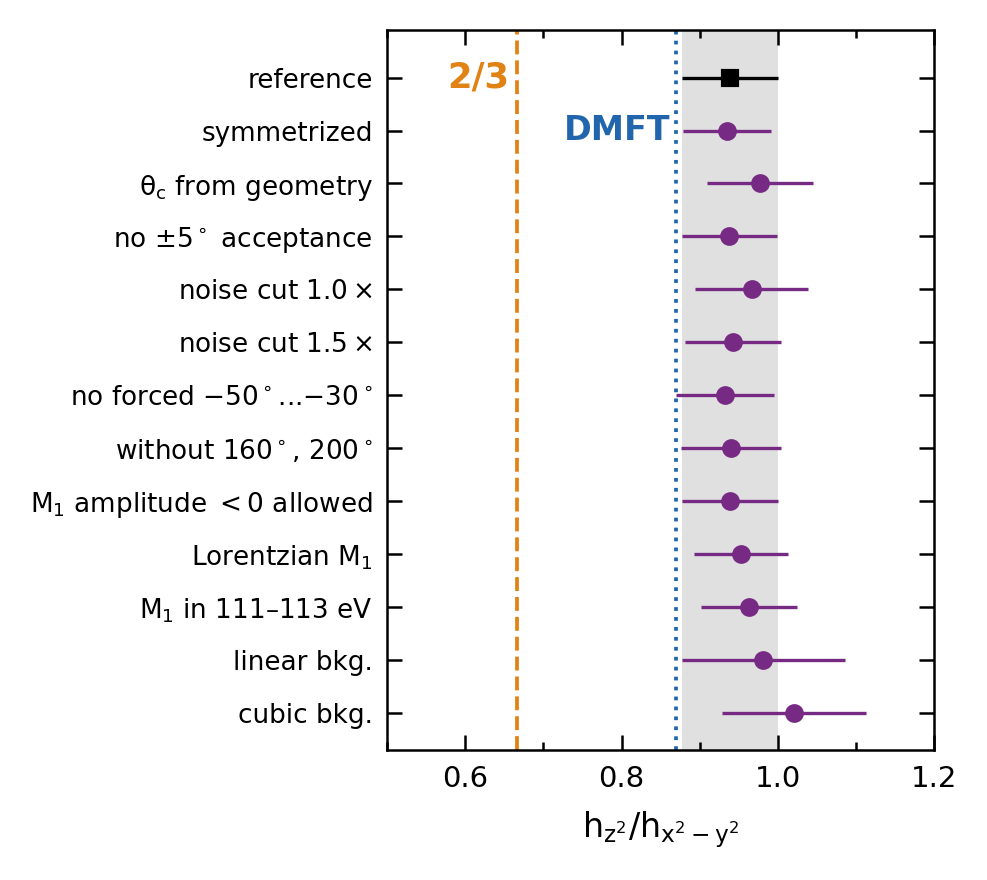}
\caption{\label{fig:variants}$\hz/\hx$ for the reference analysis (black; gray band: statistical uncertainty) and for the analysis variants of Table~\ref{tab:syst} (purple; error bars: statistical). The rows with an imposed width of the Ni peak are not shown. Dashed and dotted lines: filled bonding band (2/3) and DMFT~\cite{Shilenko2023}.}
\end{figure}

Table~\ref{tab:syst} and Fig.~\ref{fig:variants} give $\hz/\hx$ when one analysis choice is varied at a time, all other choices being those of the reference analysis. The variants concern the angular offset (fixed to the value $6.6^\circ$ expected from the x-ray diffraction orientation), the angular acceptance, the noise threshold (1.0 and 1.5 times the median), the spectra retained by hand, the amplitude of the Ni peak (negative values allowed), its profile (Lorentzian) and position (restricted to 111--113~eV), the degree of the background, and symmetrization. All variants lie between 0.932 and 1.020. Except for the background, they differ from the reference by less than 0.04. The cubic background gives the largest change, $+0.08$, with a larger uncertainty ($\pm0.09$); the linear background gives $0.98\pm0.10$ but a poor fit ($\chi^2_r=3.6$). We quote the range covered by these variants as the systematic uncertainty: $\hz/\hx=0.94\pm0.06\,(\text{stat})\,^{+0.08}_{-0.01}\,(\text{syst})$.

The ratio also depends on the width of the Ni $M_1$ peak, which is correlated with the high-energy tail of the La edge: it decreases from 1.33 to 0.83 when the width is imposed between 3 and 7~eV (last rows of Table~\ref{tab:syst}), by about $-0.11$ per eV near the fitted value. These rows are not included in the systematic range, because the width is determined by the data, $5.57\pm0.22$~eV. Imposing the limits of this statistical interval, 5.35 and 5.80~eV, gives $0.956\pm0.064$ and $0.922\pm0.059$, i.e.\ a change of $\pm0.02$ of the ratio. All widths between 3 and 7~eV give a ratio above 2/3.

\begin{table}[h]
\caption{\label{tab:syst}Hole ratio for single-parameter variations of the analysis. $N$: number of spectra used; M1 FWHM: width of the Ni peak. The last rows impose the width of the Ni peak and are not included in the systematic range.}
\begin{ruledtabular}
\begin{tabular}{lcccc}
Variant & $N$ & M1 FWHM (eV) & $h_{z^2}/h_{x^2-y^2}$ & $\chi^2_r$ \\
\hline
Reference & 23 & 5.57 & $0.939\pm0.061$ & 1.09 \\
Symmetrized reference & 23 (10 groups) & 5.57 & $0.935\pm0.056$ & 0.56 \\
$\theta_c$ fixed to the diffraction value ($6.6^\circ$) & 24 & 5.57 & $0.977\pm0.068$ & 1.35 \\
Acceptance $\pm0^\circ$ & 23 & 5.57 & $0.938\pm0.061$ & 1.07 \\
Noise cut $1.0\times$ median & 19 & 5.63 & $0.967\pm0.072$ & 1.15 \\
Noise cut $1.5\times$ median & 24 & 5.64 & $0.943\pm0.061$ & 1.09 \\
No spectra retained by hand ($-50^\circ$ to $-30^\circ$) & 20 & 5.57 & $0.932\pm0.062$ & 1.09 \\
Without $160^\circ$ and $200^\circ$ & 22 & 5.60 & $0.941\pm0.064$ & 1.13 \\
M1 amplitude of either sign & 23 & 5.57 & $0.939\pm0.061$ & 1.09 \\
Lorentzian M1 & 22 & 5.30 & $0.953\pm0.060$ & 0.79 \\
M1 position in 111--113~eV & 22 & 5.26 & $0.963\pm0.062$ & 0.89 \\
Linear background & 25 & 4.93 & $0.981\pm0.104$ & 3.59 \\
Cubic background & 24 & 5.60 & $1.020\pm0.092$ & 1.69 \\
\hline
M1 FWHM imposed to 3 eV & 22 & 3.00 & $1.327\pm0.114$ & 0.74 \\
M1 FWHM imposed to 4 eV & 24 & 4.00 & $1.131\pm0.099$ & 1.47 \\
M1 FWHM imposed to 5 eV & 24 & 5.00 & $1.023\pm0.079$ & 1.44 \\
M1 FWHM imposed to 6 eV & 23 & 6.00 & $0.909\pm0.058$ & 1.07 \\
M1 FWHM imposed to 7 eV & 22 & 7.00 & $0.825\pm0.050$ & 0.91 \\
\end{tabular}
\end{ruledtabular}
\end{table}

\section{Polar representation over $360^\circ$}
\begin{figure}[h]
\includegraphics[width=0.5\textwidth]{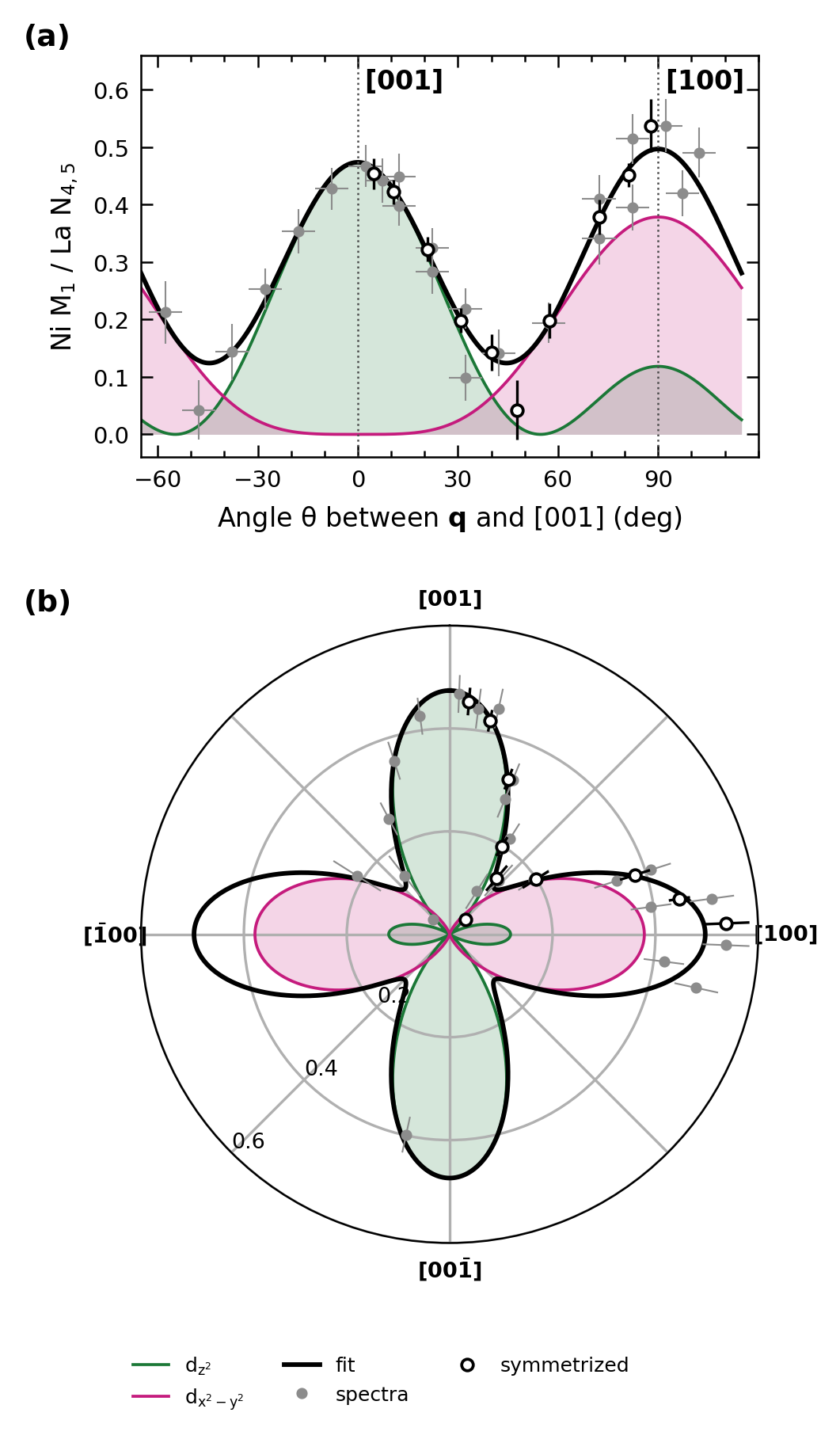}
\caption{\label{fig:polar360}Data of Fig.~3 of the main text without folding. (a) Ni $M_1$/La $N_{4,5}$ as a function of $\theta$ (as in Fig.~3a). (b) Polar representation over $360^\circ$: individual spectra (gray) at their measured angle $\theta$ between $\mathbf q$ and $z=[001]$ (the axes are labeled by their crystallographic directions, $x=[100]$); symmetry-averaged spectra (open circles) in one quadrant. Black line: fit; green and pink: $d_{z^2}$ and $d_{x^2-y^2}$ contributions.}
\end{figure}

Figure~\ref{fig:polar360} shows the data of Fig.~3 of the main text without folding. The Cartesian panel (a) is that of Fig.~3(a). In the polar panel (b), the individual spectra (gray) are drawn at their measured angle $\theta=\alpha-\theta_c$ (modulo $360^\circ$), where $\alpha$ is the nominal angle; those outside the window of Fig.~3(a) are shifted there by $180^\circ$. The symmetry-averaged spectra (open circles) are shown in a single quadrant. The intensity has the symmetry of the orbital densities, $I(\theta)=I(-\theta)=I(\theta+180^\circ)$, which is why the fit is the same in the four quadrants.

\bibliography{Biblio}